\documentclass[11pt,oneside,letterpaper]{article}
\usepackage{amssymb}
\usepackage{amsmath}
\usepackage[dvips]{graphicx}
\usepackage{setspace}
\usepackage{fancyhdr}
\usepackage{xcolor}
\usepackage{ifpdf}
\usepackage{graphicx}
\usepackage{rotating}
\usepackage{comment}

\def\p{\partial}
\def\mM{\mathcal{M}}
\def\mN{\mathcal{N}}
\def\mA{\mathcal{A}}
\def\mC{\mathcal{C}}
\def\mZ{\mathcal{Z}}

\newcommand{\ba}{\begin{array}}
\newcommand{\ea}{\end{array}}
\newcommand{\bi}{\begin{itemize}}
\newcommand{\ei}{\end{itemize}}
\newcommand{\bea}{\begin{eqnarray}}
\newcommand{\eea}{\end{eqnarray}}
\newcommand{\be}{\begin{equation}}
\newcommand{\ee}{\end{equation}}

\numberwithin{equation}{section}
\allowdisplaybreaks  

\begin{document}

\vspace*{2.5cm}
\begin{center}
{ \LARGE \textbf{Generalized Gravitational Entropy for Warped Anti-de Sitter Space }\\}
\vspace*{1.7cm}
Wei Song, Qiang Wen, and Jianfei Xu\\
\vspace*{0.6cm}
Yau Mathematical Sciences Center\\
Tsinghua University, Beijing, 100084, China\\

\vspace*{0.8cm}

\end{center}
\vspace*{1.5cm}
\begin{abstract}
\noindent
For spacetimes that are not asymptotic to anti-de Sitter Space (non AAdS), we adapt the Lewkowycz-Maldacena procedure to find the holographic entanglement entropy. The key observation, which to our knowledge is not very well appreciated, is that asymptotic boundary conditions play an essential role on extending the replica trick to the bulk.
For non AAdS, we expect the following three main modifications: (1) the expansion near the special surface has to be compatible with the asymptotic expansion; (2) periodic conditions are imposed to coordinates on the phase space with diagonalized symplectic structure, not to all fields appearing in the action; (3) evaluating the entanglement functional using the boundary term method amounts to evaluating the presymplectic structure at the special surface, where some additional exact form may contribute.
An explicit calculation is carried out for three-dimensional warped anti-de Sitter spacetime (WAdS$_3$) in a consistent truncation of string theory, the so-called S-dual dipole theory. It turns out that the generalized gravitational entropy  in WAdS$_3$ is captured by the least action of a charged particle in WAdS$_3$ space, or equivalently, by the geodesic length in an auxiliary AdS$_3$. Consequently, the bulk calculation agrees with the CFT results, providing another piece of evidence for the WAdS$_3$/CFT$_2$ correspondence.
\end{abstract}

\newpage
\setcounter{page}{1}
\pagenumbering{arabic}

\tableofcontents
\setcounter{tocdepth}{2}

\onehalfspacing

\section{Introduction.}
Entanglement plays a central role in many fields of physics, including many body systems, quantum information, and quantum field theories. In the context of AdS/CFT \cite{Maldacena:1997re,Gubser:1998bc,Witten:1998qj}, Ryu and Takayanagi \cite{Ryu:2006bv,Ryu:2006ef}  proposed (the Ryu and Takayanagi formula) that the holographic dual of the entanglement entropy is captured by the area of a minimal co-dimension 2 surface in the bulk. A covariant version (the Hubeny Rangamani and Takayanagi formula) was proposed in Ref. \cite{Hubeny:2007xt}. Large amounts of evidence \cite{Nishioka:2009un} have accumulated and an explanation as the generalized gravitational entropy was made by Lewkowycz and Maldacena \cite{Lewkowycz:2013nqa}. On the other hand, the success of holography goes beyond AdS/CFT,  for instance, the recent development of the  Kerr/CFT correspondence \cite{Guica:2008mu,Bredberg:2011hp}, flat space holography \cite{Barnich:2010eb,Hosseini:2015uba}, Schr\"{o}dinger or Lifshitz spacetime or nonrelativistic field theory duality \cite{Son:2008ye, Balasubramanian:2008dm}, etc.
Some efforts \cite{Li:2010dr,  Anninos:2013nja, Gentle:2015cfp} have been made in understanding the holographic entanglement entropy in these spacetimes without an asymptotic AdS boundary (non AAdS). However, naively using the minimal area prescription leads to some puzzles.\footnote{For example, in \cite{Anninos:2013nja}, the HRT formula was used for WAdS$_3$ without justifications, and the result only matches  the CFT expectation in the limit of a null interval being infinite. While for Lifshitz spacetime, the authors of \cite{Gentle:2015cfp} showed that the light-sheets from the holographic entanglement extremal surface corresponding to the RT or HRT formula do not reach the boundary and thus do not enclose a bulk region. Our prescription can reproduce the CFT result for any interval, which solves the puzzle in \cite{Anninos:2013nja}, and may also shed light on the similar puzzle appeared in the Lifshitz spacetime \cite{Gentle:2015cfp}.}

One of the simplest types of non AAdS spacetimes is the so-called warped AdS$_3$ spacetime (WAdS$_3$), which appears in various contexts of physics, including three-dimensional gravity \cite{Vuorio:1985ta},  extremal Kerr black holes \cite{Guica:2008mu,Bredberg:2009pv}, and cold atom systems \cite{Son:2008ye}.
It was noticed \cite{Anninos:2008fx} that the Bekenstein-Hawking entropy of WAdS$_3$ black holes can be rewritten in the form of the Cardy Formula of a CFT$_2$. Hence, it was conjectured that WAdS$_3$ is holographically dual to a CFT$_2$.\footnote{There is some evidence that the dual field theory of WAdS$_3$ with a Dirichlet type of boundary condition is the so-called dipole-CFT \cite{ElShowk:2011cm, Song:2011sr}, the IR limit of dipole deformed 2D gauge theory. The area law of the black hole entropy suggests that the  high-energy density of states of dipole-CFT is independent of the deformation parameter. See \cite{Karczmarek:2013xxa} for a discussion of holographic entanglement entropy for non-local field theories.} Based on some earlier efforts \cite{ElShowk:2011cm, Song:2011sr, Guica:2011ia,Guica:2013jza}, the boundary conditions in support of this conjecture were found \cite{Compere:2014bia}. Hereafter, we refer to this set of boundary conditions as the Dirichlet boundary conditions.
Alternatively, under the Dirichlet-Neumann boundary conditions \cite{Compere:2009zj, Compere:2013bya}, WAdS$_3$ was conjectured to be dual to the so-called warped CFT$_2$ (WCFT) featured by a Virasoro-Kac-Moody structure \cite{Hofman:2011zj}, with the evidence that the black hole entropy can also be interpreted as the Detornay Hartman and Hofman \cite{Detournay:2012pc} formula in the WCFT, an analog of the Cardy Formula.

In this Letter, we take the approach of the generalized gravitational entropy in the manner of Lewkowycz and Maldacena \cite{Lewkowycz:2013nqa}, and give a prescription of holographic entanglement entropy for non-AAdS spacetime. The key observation is that asymptotic boundary conditions play an essential role in order to extend the replica symmetry to the bulk. We expect three main modifications to the Lewkowycz-Maldacena prescription.
As a consequence, the dual of the entanglement entropy is not necessarily given by the minimal area (length) in the bulk spacetime we start with.


As an example, we  explicitly work out the holographic entanglement entropy for WAdS$_3$ with Dirichlet boundary conditions. Interestingly, the holographic entanglement entropy is given by the least action of a charged particle in WAdS$_3$, or equivalently, by the geodesic length in some auxiliary AdS$_3$. Consequently, the bulk calculation agrees with the CFT results, providing another piece of evidence for the WAdS$_3$/CFT$_2$ correspondence under the Dirichlet boundary conditions.
It will be interesting to apply our prescription to the Dirichlet-Neumann boundary conditions and find out if the bulk calculation agrees with the field theory calculation \cite{Castro:2015csg}.
It will also be very interesting to explore the range of validity of the Ryu and Takayanagi proposal. Reference \cite{Lewkowycz:2013nqa} proves it for Einstein gravity in the context of AdS/CFT.
Corrections are expected if the bulk theory is not Einstein gravity \cite{Dong:2013qoa, Castro:2014tta}. While our analysis shows that there are also corrections due to the effect of non-AAdS spacetimes.


\section{Generalized gravitational entropy for non-AAdS spacetimes.}
We adapt the Lewkowycz-Maldacena procedure \cite{Lewkowycz:2013nqa} to derive the bulk dual of entanglement entropy.
Our first assumption is the existence of holographic duality for non-AAdS spacetime, and  the compatibility of different formalisms to establish the duality. For a given non-AAdS background,  {\it consistent boundary conditions} have to be imposed in order to define the bulk gravitational theory, and furthermore to find out the holographic dual.
In the prescription of Ref. \cite{Gubser:1998bc,Witten:1998qj}, specifying the boundary condition of a bulk field
is to identify the boundary value of the field
as the source of the dual operator.
However, ambiguity in separating the source from the vev may appear \cite{vanRees:2012cw}.
As was argued in Ref. \cite{Guica:2013jza}, a proper way is to read the source and vev from the symplectic form.
Schematically, we consider the asymptotic expansion of the metric,\footnote{The explicit formula  (\ref{bc}) applies directly for WAdS$_3$. For more general non-AAdS spacetimes,   Eq. (\ref{bc}) should be replaced by the appropriate consistent asymptotic expansion.}
\bea
ds^2&=&ds_0^2+warping\,,\label{bc}\\
ds_0^2&=&\sigma^{-2}(d\sigma^2+\gamma^{(0)}_{ij}dx^i dx^j)+h_{\mu\nu}dx^\mu dx^\nu\,,\label{bct}
\eea
where $\sigma$ parameterizes the radial direction with the asymptotic boundary at $\sigma\rightarrow 0$.
$\gamma^{(0)}_{ij}$ is the source of the dual stress tensor, and therefore is identified with
the metric of the dual field theory.  $h_{\mu\nu}$ includes all subleading terms in the small $\sigma$ expansion.
The \textit{warping} terms do not source the stress tensor, but are not necessarily subleading. Note that asymptotic AdS spacetimes do not contain \textit{warping} terms.
Similarly,
the boundary expansion for all other fields, collectively denoted by $\phi_i$, gives a prescription for reading the sources $\phi_i|_{\mathrm{source}}$ that coupled to the dual operators. Conversely, to find out the bulk dual of any operator at the boundary, we have to find the bulk configuration by specifying the boundary values of the source $\phi_i|_{\mathrm{source}}$.
In many examples of holography for non AAdS spacetimes \cite{Guica:2008mu, Barnich:2010eb, Compere:2014bia}, asymptotic symmetry analysis gives some indication about the dual field theory.
In the covariant phase space formalism \cite{Barnich:2001jy,Barnich:2007bf}, there are ambiguities in deriving the presymplectic structure and symplectic structure, which will furthermore lead to some ambiguities in the definition of conserved charges.
We assume that the covariant formalism and the holographic renormalization analysis \cite{deBoer:1999tgo,deHaro:2000vlm,Skenderis:2002wp,Papadimitriou:2010as} are compatible with each other; namely, correctly fixing these ambiguities in the the covariant approach should lead to a consistent identification of the source/vev, as well as appropriate boundary terms in the holographic renormalization approach.
In the following, we will switch languages between the two formalisms.

Consider a quantum field theory on a manifold $\mathcal{N}$; the entanglement entropy can be calculated by the replica trick
\bea\label{VNEZ}
S_{EE}(\mA)&=&-n\p_n[\log \mathcal{Z}_n-n\log \mathcal{Z}_1]|_{n=1}\,,
\eea
where $\mathcal{Z}_n$ is the partition function on $\mathcal{N}_n$, which is the $n$-fold cover of $\mathcal{N}_1$, defined by first making $n$ copies of $\mathcal{N}_1\equiv \mathcal{N}$, cutting each $\mathcal{N}_1$ open at a region $\mA$, and then gluing them together cyclically. By construction, there is a $Z_n$ symmetry whose set of fixed points is the boundary of region $\mA$ denoted by $\partial \mA$.
For a field theory with Lorentzian invariance, let $\tau$ denotes the angle around $\p \mA$. Then, on $\mN_n$,
 all fields have the property $\phi^B_{i}(\tau)=\phi^B_i(\tau+2\pi)$, but with period $\tau\sim \tau+2\pi n$.
For conformal field theory, the most divergent piece of the entanglement entropy is universal \cite{Srednicki:1993im}.  Note that here we have assumed that a Euclidean theory exists, and that there is a proper way of doing Wick rotations.

Let $\mM_n$ denote the bulk extension of $\mN_n$. Then, the relation between data on $\mN_n$ and data on $\mM_n$ should be read from the asymptotic expansion (\ref{bct}). $\gamma_{ij}^{(0)}$ appearing in the bulk metric is identified with the metric on $\mN_n$,  and hence the rule of Wick rotation will be extended to the bulk.\footnote{Our prescription of the Wick rotation in the bulk is as follows.
As long as an Euclidean version of the boundary theory exists, we can use the asymptotic expansion to extend the rules of Wick rotations to the bulk.
More explicitly, when $n = 1$, we can expand the metric $ds^2_e$ in the manner as Lewkowycz-Maldacena. With the warping terms included, in the full metric $ds^2$ will in general become complex , which leads to difficulties with quantization.  In a full quantum theory, it is difficult to deal with complex saddle points in a controlled way.  However, if we assume that an Euclidean theory captures physics in Lorentzian signature and that the semi-classical limit exists, these complex metrics will still be saddle points in the path integral. Therefore the dominant contribution to the partition function will still be the on shell action on these solutions.  Furthermore, if the on-shell action is still real, the calculation of holographic entanglement entropy \`{a} la Lewkowycz-Maldacena will not be affected.  Similar discussions on complex gravity solutions also appeared in \cite{Fischetti:2014zja}, where the complex on-shell action was also considered.}
The metric of $\mM_1$ near a co-dimension 2 surface ending on $\p\mA$ can be expanded as
{\bea
ds^2&=&ds_{e}^2+tilting\,,\label{nc}\\
ds_e^2&=&dr^2+r^2 d\tau^2+(\tilde{g}_{ij}+ 2K_{aij}x^a)dy^idy^j+\mathrm{subleading}\,,\label{nct}\eea
where 
 $r$ parametrizes the separation from the curve. The \textit{tilting} terms are added to make $\tau$ the bulk extension of the circle around $\p\mA$ .
In other words, {\it the first modification to the Lewkowycz-Maldacena procedure is that the asymptotic expansion (\ref{bc}) at small $\sigma$ and the near curve expansion (\ref{nc}) at small $r$ have to be compatible with each other. }Namely, after some coordinate transformations,
the leading terms in Eqs. (\ref{bc}) and (\ref{nc}) agree with each other at the double limit $\sigma\rightarrow 0$ and $ r \rightarrow 0$; \bea
ds^2_0|_{r\rightarrow 0}&=& ds^2_e|_{\sigma\rightarrow 0}\,,\label{matchingf}\\
warping|_{r\rightarrow0} &= & tilting|_{\sigma\rightarrow 0}\,,\label{matching}\\
 \phi_i|_{source}(\tau)&=&\phi_i^B(\tau)\,.
 \eea
With Eqs. (\ref{bc}), (\ref{nc}), and (\ref{matchingf}) in mind, we can extend the replica symmetry to the bulk following the reasoning of Ref. \cite{Lewkowycz:2013nqa}.
Let $\mC_n$ denotes the set of the fixed points of $Z_n$. The expansion near $\mC_n$ is then
\bea
ds^2&=&ds_{e,n}^2+tilting\,,\label{ncn}\\
ds_{e,n}^2&=&n^2dr^2+r^2 d\tau^2+(\tilde{g}_{ij}+2K^{(n)}_{aij}x^a)dy^idy^j+\mathrm{subleading}\,.\label{nctn}
\eea
In order to select a special co-dimension 2 surface, we need to impose some regularity conditions of the fields near $\mC_n$. In Ref. \cite{Lewkowycz:2013nqa}, the requirement is that all fields and their variations are periodic. However, this condition could be overdetermining when there are mixings between various fields, as in our example later. As we mentioned before, the matching between the bulk and the boundary is through the source/vev relation, or equivalently, through the symplectic structure. Note that in the covariant formalism, the symplectic structure can be defined in the bulk \cite{Compere:2014bia}. Therefore, {\it we propose the second modification to the Lewkowycz-Maldacena procedure: to impose a periodicity condition for all the independent coordinates on the phase space, and their variations}. In particular,
\be \delta \phi_i(\tau)=\delta \phi_i(\tau+2\pi)\,,\label{ncbc}
\ee
for variations appearing in the symplectic form $\omega[\phi,\delta_1 \phi, \delta_2\phi]$.
Assuming analytic continuation,
and plugging (\ref{ncn}) and (\ref{ncbc}) into the equations of motion (EoMs) will determine the shape of a special surface $\gamma_{\mA}\equiv \mC_n|_{n\rightarrow 1}$.

We will use the boundary term method \cite{Lewkowycz:2013nqa} to evaluate the entanglement entropy (\ref{VNEZ}).\footnote{Additional terms may appear for higher derivative gravity \cite{Dong:2013qoa}, we will only consider theories without this complication.} At the classical level, $\mZ_n= \exp(-S_{\mathrm{rn}}[\mM_n])$ where $S_{\mathrm{rn}}[\mM_n]$ is the renormalized Euclidean action on a bulk manifold $\mM_n$ with replica symmetry and the set of fixed points $\mC_n$. In the sense of (\ref{bc}), the boundaries of $\mM_n$ and $\mC_n$ are $\mN_n$ and $\p\mA$, respectively. Replica symmetry requires that
 $S_{\mathrm{rn}}[\mM_n]=nS_{\mathrm{rn}}[\hat{\mM}_n]$,
where the orbifold $\hat{\mM}_n\equiv\mM_n/Z_n$
has an asymptotic boundary $\mN_1$
and a conical defect with opening angle $2\pi/n$ at a co-dimension 2 surface $\mC_n$.
As was discussed in Ref. \cite{Lewkowycz:2013nqa}, the conical defect will not contribute to $S_{rn}[\hat{\mM}_n]$}. At the classical level, the entanglement entropy can then be calculated by
\begin{align}
S_{EE}=\partial_nS_{\mathrm{rn}}[\hat{\mathcal{M}}_n]\big|_{n\to 1}=\partial_nS_{E}[\hat{\mathcal{M}}_n]\big|_{n\to 1}\,,
\end{align}
where $S_E$ is the bulk Euclidean action.
The second equality above is due to the compatibility between the asymptotic expansion (\ref{bc}) and the near cone expansion (\ref{nc}), which guarantees that the asymptotic boundary terms cancel out.
For small $n-1$, we see the difference between the metric Eq. (\ref{ncn}) and Eq. (\ref{nc}) is of order $n-1$, and therefore $S_{EE}={\delta_n S_{E}\over n-1}|_{n\rightarrow 1}.$

The variation of the Lagrangian  $\mathbf{L}$ can be written in the following form
\begin{align}\label{varL}
 \delta\mathbf{L}=\sum_{i} \mathbf{E_{\phi_i}}\delta\phi_i+\mathbf{d}\mathbf{\Theta}(\phi_i,\delta\phi_i)\,,
 \end{align}
where the $\phi_i$ represent all the metric and matter fields, and the $\mathbf{E_{\phi_i}}$ are their corresponding EoMs. The presymplectic form $\mathbf{\Theta}$ is only defined up to the addition of an exact form,
$\mathbf{\Theta}\to \mathbf{\Theta}+\mathbf{d Y}(\phi_i,\delta\phi_i)\,.$
Different choices of $\mathbf{Y}$ will affect asymptotic charges, and thus affect how holography works, see Ref. \cite{Compere:2014bia} for more discussion. Thus, {\it we propose the third main modification to the Lewkowycz-Maldacena procedure: the entanglement entropy can be calculated using the presymplectic structure, which is subject to some ambiguity. The ambiguity is fixed by requiring holography to work in the correct way.}  More explicitly,
\begin{align}\label{See1}
S_{EE}=-\int ~ \frac{\mathbf{d\Theta}(\phi_i,\delta_n\phi_i)}{n-1}\Big|_{{n\to 1}}=\int _{\gamma_{\mA}\times S^1} {\bf\Theta }(\phi_i,\p_n\phi_i)\Big|_{{n\to 1,r\to 0}}\,,
\end{align}
which is just a surface integral on $\gamma_\mathcal{A}$ after we integrate out $\tau$ along the $\tau$ circle $S^1$.

Note that most of the discussions are general, although we used some explicit expressions for some special examples in order to illustrate the idea.

\section{Review of WAdS$_3$ in the S-dual dipole truncation.}\label{section C}
In this section, we review the asymptotic symmetry analysis of WAdS$_3$ in the three-dimensional S-dual dipole truncation \cite{Detournay:2012dz} (see also Ref. \cite{Colgain:2010rg} for other consistent truncations)
\begin{align}\label{Lag}
S=\frac{1}{16\pi G_3}&\int d^3 x \sqrt{-g}\Big[R-4(\partial U)^2-\frac{4}{\ell^2}e^{-4U}A^2
\\\nonumber
&
+\frac{2}{\ell^2}e^{-4U}(2-e^{-4 U})-\frac{1}{\ell}\epsilon^{\mu\nu\rho}A_{\mu}F_{\nu\rho}\Big]\,,
\end{align}
 a consistent truncation of type  \uppercase\expandafter{\romannumeral2}B supergravity. WAdS$_3$ is a classical solution with constant $U$. 
Following Ref. \cite{Lewkowycz:2013nqa}, we assume that the hairless classical solution with replica symmetry is the dominant contribution to the Renyi entropy around $n=1$, which means that all the propagating modes are turned off.\footnote{In the main text we only focus on the sector with fixed $U$, thus all the propagating modes are turned off. In fact, some of these propagating modes discussed in section 4.1 of \cite{Compere:2014bia} may become dangerous tachyons in the large $n$ limit, as was noticed in \cite {Belin:2013dva}. However, it was also argued  in \cite {Belin:2013dva} that with Dirichlet boundary conditions, the dominant saddle is still hair-less solution around $n=1$. Therefore we will only focus on the sector with U fixed and the entanglement entropy will not be affected. Under this situation, all perturbations can be  locally written as diffeomorphisms, and therefore $\delta_n U=0.$ On the other hand, non-constant perturbations of $U$ will lead to quantum corrections to the entanglement entropy, and is expected to distinguish WAdS$_3$ from AdS$_3$. We hope to report the progress in the future.} In the following, we will only consider the classical contribution, and will focus on the sector with fixed $U$.
In terms of an auxiliary metric $\tilde{g}_{\mu\nu}$ and field $\tilde{A}_{\mu}$ given by
\bea\label{bridge}
\tilde{g}_{\mu\nu}&=&e^{-4U}g_{\mu\nu}+A_{\mu}A_{\nu}\,,\\
\tilde{A}_{\mu}&=&A_{\mu},  \tilde{A}^{\mu}=\tilde{g}^{\mu\nu}\tilde{A}_{\nu}=A^{\mu}\,,\eea
the EoMs of the sector with fixed $U$ can be written as
\bea
\label{eq3}
&&\tilde{R}_{\mu\nu}+\frac{2}{\ell^2}\tilde{g}_{\mu\nu}=0\,,\\
&&\tilde{F}_{\mu\nu}=\frac{2}{\ell}\tilde{\epsilon}_{\mu\nu\lambda}\tilde{A}^{\lambda}\,,
\,
\tilde{A}^{2}=1-e^{-4U}\,,\label{eq4}
\eea
with ${A}$ being a Killing vector in both $\tilde{g}$ and $g$,
\bea\label{eq5}
\nabla _{(\mu}A_{\nu)}=
\tilde{\nabla}_{(\mu}\tilde{A}_{\nu)}=0\,.\eea
Equation (\ref{eq3}) is just the EoMs of the three-dimensional Einstein gravity with a negative cosmological constant $-{1/\ell^2}$, which means that $\tilde{g}$ is locally AdS$_3$.
Equations (\ref{eq4}) and (\ref{eq5}) define a self-dual Killing vector $\tilde{A}$ with constant norm.
On the other hand,
given a locally AdS$_3$ solution $\tilde{g}$,  and a self-dual Killing vector $\tilde{A}$ with constant norm, then (\ref{bridge}) uniquely defines a solution of S-dual dipole theory. In particular, the metric $g_{\mu\nu}=e^{4U}(\tilde{g}_{\mu\nu}-\tilde{A}_\mu\tilde{A}_\nu)$ is then locally WAdS$_3$ by definition.

A Dirichlet type of boundary conditions in support of the WAdS$_3$/CFT$_2$ conjecture \cite{Anninos:2008fx} was found in Ref. \cite{Compere:2014bia}.
In the sector with fixed $U$, the boundary conditions can be written as
\bea
\label{bcwads}{ds^2\over\ell^2}&=&(1-\chi)^{-1}\big(d{\tilde s}^2-A^2\big)\,,\\
{{A}\over\ell}&=&\chi e^\Phi dt^+-\Big({e^{-\Phi}\over2}\sigma^{-2}+{\chi e^\Phi L \over 2k }\sigma^2\Big)dt^-\,,\\
{d\tilde{s}^2\over\ell^2}&=&{d\sigma^2-dt^+dt^-\over\sigma^2}-{\chi e^{2\Phi}L\over k }\sigma^2dt^+dt^-+{L\over k}(dt^-)^2
\nonumber\\
&&
+\Big({1\over4}(\p_+\Phi)^2+\chi e^{2\Phi}\Big)(dt^+)^2+{\p_+\Phi\over\sigma}d\sigma dt^+,\label{bcads}\eea
where $\chi=1-e^{-4U}$ is a constant, $\Phi=\Phi(t^+)$, and $L=L(t^-)$.
For the WAdS$_3$ black string solution, both $\Phi$ and $L$ are constant.
Note that this asymptotic expansion terminates, and Eq. (\ref{bcwads}) is the full nonlinear solution of the S-dual dipole theory with $U$ fixed.
The dual
field theory is then defined on a fixed two-dimensional Minkowski space \be ds^2_{boundary}\equiv \gamma^{(0)}_{ij}dx^idx^j=-dt^+dt^-\rightarrow dzd{\bar z}\,,\ee
where
$t^+\rightarrow z, t^-\rightarrow -{\bar z}$ bring the solution to the Euclidean signature.
Note that Eq. (\ref{bcwads}) is of the form of Eq. (\ref{bc}), with the first term coming from the expansion of the auxiliary metric $d\tilde{s}^2$, and \be warping=-e^{4U}A^2|_{\sigma\rightarrow0}\,.\ee
 As was argued in Ref. \cite{Compere:2014bia},
 the asymptotic symmetry group is generated by left and right moving Virasoro generators if we fix the ambiguity in the presymplectic structure  by choosing
 \begin{align}\label{Y}
\mathbf{Y}=-{1\over16 \pi G_3}\epsilon_{\mu\alpha\beta}A^{\alpha}\delta A^{\beta}dx^{\mu}\,,
\end{align}
then, the presymplectic structure can be written as
$\mathbf{\Theta}=\frac{1}{16\pi G_3}\left(\mathbf{\Theta}_{g}+\mathbf{\Theta}_{A}+\mathbf{\Theta}_{\mathbf{Y}}\right)\,,$
 where $\mathbf{\Theta}_{g}$ and $\mathbf{\Theta}_{A}$ are the boundary terms that can be read directly from the variation of the action in the gravity sector and vector sector respectively, and $ \mathbf{\Theta}_{\mathbf{Y}}=16 \pi G_3\bf{dY}$. There are no contributions  from the scalar sector for constant $U$. With the choice of Eq. (\ref{Y}), the S-dual dipole theory on WAdS$_3$ has the same symplectic structure as that of Einstein gravity on AdS$_3$ defined by Eq. (\ref{bridge})\be
\mathbf{\Theta}=\tilde{\mathbf{\Theta}}\,.\label{presym}\ee

With the boundary condition (\ref{bcwads}), and the choice of the presymplectic structure (\ref{Y}), a standard analysis in the manner of Brown-Henneaux \cite{Brown:1986nw} shows that the asymptotic symmetry group of WAdS$_3$ in the constant $U$ sector of Eq. (\ref{Lag}) is generated by two sets of Virasoro generators, which indicates that the holographic dual is a CFT$_2$  at the semiclassical approximation.

\section{Generalized gravitational entropy for WAdS$_3$.}\label{section D}
Now, we try to find the holographic entanglement entropy for WAdS$_3$.
At this point,
we assume a standard replica trick for the CFT, and only consider the classical contributions.
We need to write down a near curve ansatz that is compatible with the boundary condition (\ref{bcwads}).  The AdS$_3$ metric $d\tilde{s}^2$ with Dirichlet boundary conditions can be written in the form of Eq. (\ref{nctn}).
Because of relation (\ref{bridge}),  a natural ansatz for $ds^2$ is then to set
\be tilting=- e^{4U}{ A}^2|_{r\rightarrow 0}\,. \ee
This choice guarantees that the criterion (\ref{matching}) is satisfied.
Effectively, this means that we expand $g$ by expanding the auxiliary metric $\tilde{g}$ in the form of Eq. (\ref{nct}) without tilting:
\begin{align}
ds^2=&e^{4U}(d\tilde{s}^2-A^2)\\
 d\tilde{s}^2=&n^2dr^2+r^2 d\tau^2+(\tilde{g}_{yy}+2K^{(n)}_{ayy}x^a)dy^2+subleading\,. \label{stnc}
\end{align}
For $n=1,$ there is a local coordinate transformation between Eqs. (\ref{stnc}) and (\ref{bcads}), which determines the expansion of $A$ as well.
For small $n-1$, $\delta_n g$ and $\delta_n A$ should satisfy the linearized equations, which means that $\delta_n {\tilde g}$ and $\delta_n \tilde{A}$ satisfy the linearized form of Eqs. (\ref{eq3}) and (\ref{eq4}). Note that at the fixed $U$ sector, $\tilde{A}$ is not dynamical, and hence does not correspond to a source. Thus we do not set any periodicity condition to $\tilde A$ independently. Note that, this will not break the replica symmetry of the on-shell action $S_{\mathrm{rn}}[\mM_n]$.  On the other hand, $\delta_n \tilde{g}$ appears in the symplectic form, which means that we should impose the periodicity condition to the trace of the extrinsic curvature
\be K_a(\tau)=K_a(\tau+2\pi), \quad \delta_n K_a(\tau)=\delta_n K_a(\tau+2\pi)\,,\label{rg}\ee
where $K_a=\tilde{g}^{yy}K_{ayy}$. Plugging Eqs. (\ref{stnc}) and (\ref{rg}) into the linearized EoMs (\ref{eq3}), we find $K_a=0$.
This indicates that the curve selected by replica symmetry is a geodesic in the auxiliary AdS$_3$ spacetime with metric $\tilde{g}$, but not necessarily a geodesic in the original WAdS$_3$.
Furthermore, using Eqs. (\ref{See1}) and (\ref{presym}) we get
\begin{align}\label{See3}
S_{EE}&=\int^{2\pi}_{0} d\tau\int dy \sqrt{\tilde{g}}~ \tilde{\Theta}^{r}(\phi_i,\partial_n\phi_i)\Big|_{{n\to 1,r\to 0}}
\cr
&=\frac{1}{4G_3}\int_{\gamma_{\mathcal{A}}} \sqrt{\tilde{g}_{yy}}dy=\frac{\widetilde{\mathrm{Length}}(\gamma_\mathcal{A})}{4G_3}\,,
\end{align}
where $\tilde{\Theta}^{\mu}=-\frac{1}{2}\tilde{\epsilon}^{\mu\nu\rho}\tilde{\Theta}_{\nu\rho}$
and $\widetilde{\mathrm{Length}}(\gamma_A)$ denotes the length of $\gamma_A$ calculated with the metric $\tilde{g}$.
We see that the holographic entanglement entropy is the geodesic length using the auxiliary metric $\tilde{g}$. The direct calculation of $S_{EE}$ in the original WAdS$_3$ (see appendix) gives the same result.

For a WAdS$_3$ parameterized by $\chi, \Phi\,,\bar{L}$, the auxiliary metric  $\tilde{g}$ is a  BTZ black hole with temperature $T_+=\sqrt{\chi}e^\Phi\,,T_-=\sqrt{ L/k}$.
Therefore the holographic entanglement entropy of region $\mA$ can be written as
 \be S_{EE}(T_+, T_-; \mA) ={c\over6}\log \left(\frac{\sinh(T_+\Delta t^+)}{\varepsilon T_+}\frac{\sinh(T_-\Delta t^-)}{\varepsilon T_-}\right)\,,\ee
where $\epsilon=\sigma_0$ is the UV cutoff. This result agrees with the CFT expectation.

We can also evaluate the entropy formula (\ref{See3}) at the horizon. It is easy to check that a spacelike geodesic at the horizon of WAdS$_3$ black string is also a geodesic of the auxiliary AdS$_3$ black string, and (\ref{See3}) gives the same result as HRT. Therefore, our result is compatible with the intuition that the area law is universal, and that the dipole deformation is an irrelevant deformation.

\section{World-line action as the holographic entanglement entropy.}
\label{section E}
Although the curve $\gamma_\mathcal{A}$ is not necessarily a geodesic in the WAdS$_3$, we find it is actually the trajectory of a charged particle moving in the WAdS$_3$. Furthermore, in WAdS$_3$, the holographic entanglement entropy $S_{EE}$ is
given by the least action of this charged particle.
The geodesic equation in the AdS$_3$ can be written in terms of the original WAdS$_3$ as
\bea\label{eomgamma1}
&&{\ddot x}^{\mu}+\Gamma^{\mu}_{\alpha\beta}{\dot{x}^{\alpha}}{\dot{x}^{\beta}}={q\over m}{F}^{\mu}{}_{\nu}{\dot {x}^{\nu}\,,}\\
&& \quad  g_{\mu\nu} {\dot x^{\mu}}{\dot x^\nu}=1, \quad {q\over m}\equiv  A_\mu \dot{x}^\mu e^{4U} \,,
\eea
where we have used the fact that $A$ is a Killing vector, and the dot denotes derivative with respect to the affine parameter along the curve.
Also, Eq. (\ref{eomgamma1}) can be derived from the world-line action of a particle with mass $m$ and charge $q$
\bea
\label{sk}S_{m,q}&=&\frac{m}{4G_3}\int  \sqrt{g_{\mu\nu}d{x}^{\mu}d{x}^{\nu}}+\frac{q}{4G_3}\int { A}
\eea
with the gauge choice and additional constraint \be g_{\mu\nu} {\dot x^{\mu}}{\dot x^\nu}=1, \quad {q\over m}= {A_\mu}  {\dot x^{\mu}}e^{4U}\,,\label{ak}\ee
which fixes the reparametrization symmetry and the conserved momentum  along  $A^\mu$ respectively. Solving the system  (\ref{sk} and \ref{ak}) with one end point $p_1$ on the boundary will determine the other end point $p_2$.\footnote{To consider multi-intervals, it seems that we have to include more charged particles, which we will leave for future investigations.}

Furthermore, with the choice \be m=e^{-2U}\big(1+e^{-4U}({q\over m})^2\big)^{-{1\over2}}\,,\ee
the on-shell action $S_{m,q}$ on this solution calculates the entanglement entropy of the an interval $\mA$ with the two end points $p_1$ and $p_2$, ie $S_{m,q}=S_{EE}(\mA)\,.$

Similar discussions appear in Ref. \cite{Castro:2015csg}, where the entanglement entropy of a single interval in the WCFT is holographically associated with the world-line action of a charged particle in lower spin gravity.

\section*{Acknowledgements}
We thank D. Anninos, A. Castro, G. Compere, X. Dong, M. Guica, J. Maldacena, T. Takayanagi, and J. Wu for insightful discussions.  This work was supported in part by start-up funding Grant No. 543310007 from Tsinghua University. W.S. is also supported by the National Thousand-Young-Talents Program of China.



\appendix
\setcounter{equation}{0}  
\section{Direct calculation using $g$}  
\label{section:appA}
In this section, we calculate the generalized gravitational entropy $S_{EE}$ in WAdS$_3$ directly.
We choose the near curve coordinates $r,\tau$ and $y$. The metric in Warped AdS$_3$ can be decomposed into
\bea
ds^2&=&e^{4U}(d\tilde{s}^2-A^2)\\
 d\tilde{s}^2&=&n^2dr^2+r^2 d\tau^2+(\tilde{g}_{yy}+2K^{(n)}_{ayy}x^a)dy^2+subleading\,. \label{stnca}\
\eea
The regularity conditions are
\be K_a(\tau)=K_{a}(\tau+2\pi),\qquad \delta_n K_a(\tau)=\delta_n K_{a}(\tau+2\pi)\,.\label{rga}\ee
The EoMs of the sector with fixed U can be written as (Eq. (18)-(20) in the main text)
\bea
\label{eq1}
&&\tilde{R}_{\mu\nu}+\frac{2}{\ell^2}\tilde{g}_{\mu\nu}=0\,,
\\
\label{eq2}
&&\tilde{F}_{\mu\nu}=\frac{2}{\ell}\tilde{\epsilon}_{\mu\nu\lambda}\tilde{A}^{\lambda}\,,
\qquad
\tilde{A}^{2}=1-e^{-4U}\,,
\qquad
\nabla _{(\mu}A_{\nu)}=
\tilde{\nabla}_{(\mu}\tilde{A}_{\nu)}=0\,.
\eea
Plugging the anstz (\ref{stnca}) and regularity condition (\ref{rga}) into (\ref{eq1}), the $n$-th solution with a conical defect in $d\tilde{s}^2$ is given by
\bea\label{hatg}
d\tilde{s}^2&=&\left[n^2+\mathcal{O}( r^2)\right]dr^2 +\left[r^2+\mathcal{O}( r^4)\right]dt^2
 +\mathcal{O}( r^2)~dt dy+\mathcal{O}( r^2) ~dr dy\cr
&&+\left[f(y,n)+\mathcal{O}( r^2)\right]dy^2\,.
\eea
  As a gauge choice, we set $f(y,1)=1$. A general ansatz for the vector field ${A}_{\mu}$ with finite norm can be written as
\begin{align}\label{hata}
A= r a_{\tau}(y,\tau,r,n)d\tau+a_r(y,\tau,r,n)dr+a_y(y,\tau,r,n)dy\,.
\end{align}
Plugging into (\ref{eq2}), we get
\begin{eqnarray}\label{as}
\partial_\tau a_y(y,\tau,0,n)&=&0 \,,
\\
a_y(y,\tau,0,n)^2&=&1-\alpha (y,n)^2-\beta (y,n)^2-e^{-4 U}\,,
\nonumber\\
a_\tau(y,\tau,0,n)&=&\cos \left(\frac{\tau}{n}\right) \alpha (y,n)+\sin \left(\frac{\tau}{n}\right) \beta (y,n)\,,
\nonumber\\
a_r(y,\tau,0,n)&=&n \big[\sin \left(\frac{\tau}{n}\right) \alpha (y,n)-\cos \left(\frac{\tau}{n}\right) \beta (y,n)\big]\,.\nonumber
\end{eqnarray}
Note that  $\delta_n A_\mu(\tau)\neq \delta_n A_\mu(\tau+2\pi)$.  As we argued in the main text, we do not impose periodicity condition on $\delta_n A$, since it does not show up in the symplectic structure. As we will see more explicitly later,  the on-shell action is independent of $\tau$, which makes sure that $S_{E}[\mM_n]=nS_{E}[\hat{\mM}_n]$.
Plugging (\ref{stnca}) and (\ref{as}) into the holographic entanglement entropy
$
S_{EE}=\partial_nS_{E}[\hat{\mathcal{M}}_n]\big|_{n\to 1}
$, it is straight forward to see that the contribution only comes from boundary terms, which is captured by the presymplectic structure \cite{Compere:2014bia},
\bea\label{theta}\Theta^{\mu}_{g}&=&g^{\mu\nu}g^{\rho\sigma}\nabla_{\rho}\delta_ng_{\sigma\nu}-g^{\mu\nu}g^{\rho\sigma}\nabla_{\nu}\delta_ng_{\rho\sigma}\,,
\\
\Theta^{\mu}_{A}&=&-\frac{2}{\ell}\epsilon^{\mu\nu\rho}A_{\rho}\delta_{n}A_{\nu}\,,\nonumber
\\
\Theta^{\mu}_{Y}&=&-\nabla_{\nu}(A^{\mu}\delta_{n}A^{\nu}-A^{\nu}\delta_{n}A^{\mu})\,.\nonumber
\eea
Plugging (\ref{stnca}) and (\ref{as}) into (\ref{theta}) we get
\begin{align}
& \int^{2\pi}_{0}d\tau \sqrt{g}~\Theta^{r}_{A}=0\,,
\\
& \int^{2\pi}_{0}d\tau \sqrt{g}~\Theta^{r}_{Y}=-2 \pi  e^{4 U} \left[\alpha (y,1)^2+\beta (y,1)^2\right]\,,
\\
& \int^{2\pi}_{0}d\tau \sqrt{g}~\Theta^{r}_{g}=2 \pi  \left[e^{4 U} \left[\alpha (y,1)^2+\beta (y,1)^2\right]+2\right]\,.
\end{align}
It is easy to see that, the undetermined functions $\alpha(y,1)$ and $\beta(y,1)$ will cancel with each other when we add up all these three sectors. So the holographic entanglement entropy is then given by
\begin{align}\label{See4}
S_{EE}=\frac{1}{4 G_3}\int dy=\frac{\widetilde {Length}({\gamma}_\mathcal{A})}{4G_3}\,,
\end{align}
which is just the result (Eq. (32) in the main text) we calculated from the auxiliary AdS$_3$. In general $g_{yy}|_{n\to 1, r\to 0}$ is not $1$ under this gauge, therefore $\widetilde {Length}({\gamma}_\mathcal{A})$ is not the length of $\gamma_A$ evaluated using $g$.


\end{document}